\documentstyle[12pt]{article}
\title{Perturbative Static Four-Quark Potentials}
\author{J.T.A.Lang\thanks{E-mail address:
lang@lswes7.ls-wess.physik.uni-muenchen.de},
J.E.Paton\thanks{E-mail address: paton@thphys.ox.ac.uk},\\
Department of Physics (Theoretical Physics) \\
1 Keble Road, Oxford OX1 3NP, UK \\
A.M. Green\thanks{E-mail address: green@phcu.helsinki.fi}\\
Research Institute for Theoretical Physics,\\  University
of Helsinki, Finland}

\begin{document}
\maketitle
\begin{abstract}
A first attempt to understand hadron dynamics at low energies
in terms of the fundamental quark and gluon degrees of freedom
incorporates the effects of the gluonic field into a potential
depending only on the spatial positions of the quarks,
which are considered in the infinite mass limit.
A suitable framework for calculating such potentials
between static quarks, i.e.\ a generalization of the Wilson loop
will be discussed.

Making a connection with recent Monte Carlo lattice simulations
for the lowest two energies of a system of two quarks
and two antiquarks, the static $qq\bar{q}\bar{q}$-potential
will be calculated in perturbation theory to fourth order.
The result will be shown to be exactly equal to the prediction of a
straightforward two-body approach, which in Monte Carlo lattice
simulations has been found to be a reasonable approximation
for very small interquark distances.
\end{abstract}

\section{Introduction}
As QCD is the accepted theory of the strong interactions, it is no doubt
desirable to understand all hadronic phenomena directly in terms of the
fundamental fields of QCD.
However, QCD being asymptotically free, perturbation theory is
applicable only for very short distances and cannot cover the complete range
of interest. At present lattice gauge simulations are the only way to study
such systems. In a first approach, the static approximation is the natural
choice,
where the gluonic degrees of freedom are integrated out, and quark loops are
ignored (the quenched approximation), giving rise to a potential between the
stationary quarks. The potential of the quark-antiquark system, where this
approach---leading to the familiar Wilson Loop---is very well known,
has been calculated extensively in Monte Carlo lattice simulations.
(For recent data, see e.g.~\cite{booth}.)
The ground state potential of this static system has also been calculated
in perturbation theory upto sixth order~\cite{fisch}.

Here we shall describe how to generalize this procedure to multi-quark systems,
especially to $(q\bar{q})^k$ systems. However, even
with present-day computers, $q\bar{q}$ lattice simulations are
still very demanding, and the amount of computations needed increases rapidly
with the number of interacting quarks. Reliable models for multi-quark systems
expressing their
potentials e.g.\ in terms of the well known $q\bar{q}$-systems
would therefore be of great help. Such two-body approximations have proven
successful in many areas of physics, and these  models can be formulated
without difficulty. For the $qq\bar{q}\bar{q}$-system, which is the simplest
one that can be considered consisting of two colour singlets, this model has
been tested against numerical data
from a Monte Carlo simulation \cite{paton}. For
small distances the agreement has been found reasonable. It has also been
observed \cite{mor} that the two-body model corresponds
to lowest order perturbation theory.
We shall be able to show that it is correct even to fourth order. To
sixth order, however, three- and four-body forces begin to appear.

\section{The Generalized Wilson Loop}
While the concepts discussed below
are of course well known in the context of the
Wilson Loop for the $q\bar{q}$-system,
we find it useful to start with rephrasing these concepts in the case of an
arbitrary number
of quarks, leading to a study of more complicated systems.

When we have assembled a
system of several quarks (and antiquarks),
gluons will mediate a force between them.
Treating this system in an approximation
as a quantum mechanical system of several static
quarks, the interactions between the quarks
are incorporated into a potential.
This assembly of quarks is then expected to propagate
in time with the usual factor of $e^{-itH}$,
where the interesting piece of the Hamilton operator
$H$ is the potential energy.
Thus, by calculating appropriate Green functions,
the potentials of eigenstates
of $H$ can be extracted.

\subsection{Setting up Gauge Invariant States}
\label{sin}
Because of confinement, it makes sense only to talk about systems of quarks
where the overall
states have colour singlet quantum numbers.
The problem with setting up say a
$q\bar{q}$-system in a singlet is that the quark and antiquark
are located a distance apart.
This problem can be overcome by inserting the path ordered exponential
$U(x,y,A) = {\cal P}e^{ig\int^x_y T^aA_a^{\mu}(z)dz_{\mu}}$ between the
locations
$x$ and $y$ of the quarks in the presence of the gauge potential $A$.
Here $g$ denotes the coupling constant and $T^a$ the representation matrices.
Thus $\bar{\psi}(x)U(x,y)\psi(y)|0\rangle$ will serve as a basis state
in this case.
We must also know how many basis states there are. When dealing with Green
functions coming from Monte Carlo lattice simulations,
they will have contributions from excited states of the gluonic field,
and there are infinitely many of them even in the $q\bar{q}$-case.
With suitable methods, the lowest potentials can be extracted, and several
have been calculated for the quenched $q\bar{q}$-system -- see for example
\cite{hunt}, \cite{mich}.
The situation is different for Green functions calculated in perturbation
theory. Here, unlike the lattice simulations,
we can and must go to the infinite time limit.
We do not expect to reach excited states of the gluon field in finite order
perturbation theory, and thus the number of basis states for a system
of several quarks is given by the usual arguments of group representation
theory, e.g.\ one for the
$q\bar{q}$-system and two for the $qq\bar{q}\bar{q}$-system.
In the large time limit we expect that the effects of `introducing'
the quarks into the vacuum will be irrelevant in comparison
to their time evolution, and the notion of a potential makes sense.

\subsection{Diagonalization}

\label{diag}
It may be shown that the state
\[
|\mbox{ quarks }q_i\mbox{ and antiquarks }\bar{q}_j\mbox{ at time
}t\rangle
\]
\begin{equation}
= \bar{\psi}_{i}(-t,x_i)\ldots U(-t,x_i,y_i,A)\ldots \psi(-t,y_i)|0\rangle
\label{e1}
\end{equation}
satisfies Schr\"odinger's equation.
Forming the overlap of states at time $-t$ and $t$,
we get an equation between Green functions
and expressions of the form
${\bf A}_{ij}\stackrel{def}{=}\langle A_i|e^{-itH}|A_j\rangle$,
where $|A_i\rangle$ stands for some basis state and we have introduced the
matrix $\bf A$.
By assuming a decomposition of these basis states into eigenstates of $H$,
a diagonalization procedure will yield the potentials.
In the case of the Green functions coming from lattice
simulations, one considers a practical number of basis states,
expands them in energy eigenstates and drops contributions
with $e^{-itE_i}$ for energies $E_i$ above a certain limit.
Of course we implicitly assume Wick-rotation.
In perturbation theory, where a power expansion of $e^{-itE_i(g)}$ in
the coupling
$g$ will not be exponentially damped, we need to consider all linearly
independent basis
states, a number that is finite, as remarked in the last section.
Because of this finiteness,
we can find an invertible transformation to energy eigenstates,
and the diagonalization is straightforward.
In fact, given a matrix ${\bf A}$ satisfying certain
consistency relations, we can perturbatively
prove \cite{t} the existence
of a time-independent basis transformation
such that in this new basis ${\bf A}$ is not only diagonal,
but its eigenvalues are of the form $e^{-itE_i(g)}$.
Here the energy $E_i(g)$ of
the $i$-th
basis state, which can be calculated perturbatively, is for static quarks
equal to the $i$-th potential (apart from an irrelevant constant, the rest
mass).

\subsection{Loops}
\label{loops}
What remains to be done is to bring the Green functions
of the last paragraph to more familiar forms.
Since we work within the static approximation, the full quark
propagator in the presence of gauge fields can be
calculated \cite{eichtenfeinberg}:
\begin{eqnarray}
S_0(x,y,A)&=&-i[{\cal P}e^{ig\int^x_yT^aA_a^{\mu}(z)dz_{\mu}}]
e^{-im|x^0-y^0|}\delta(\vec{x}-\vec{y})
\times\nonumber\\
&&[\frac{1+\gamma^0}{2}\Theta(x^0-y^0)+\frac{1-\gamma^0}{2}
\Theta(y^0-x^0)]
\label{eq21}
\end{eqnarray}

We shall now outline how various contour integrations, i.e.\ loops arise.
Considering the
well-known $q\bar{q}$-case, we find a path-ordered
line integral from antiquark to quark
arising from the $U$ in eq.~(\ref{e1}), then the path-ordered line integral
propagating
the quark forward in time from eq.~(\ref{eq21}).
Another $U$ and the antiquark
propagating
backwards in time close the rectangle of the familiar Wilson loop.
Starting with the Green functions described below eq.~(\ref{e1})
and evaluating them for propagation from $-t/2$ to $t/2$,
the following diagrammatic rule for calculating the Green function
dealing with an arbitrary number $k/2$ of quark-antiquark pairs
(i.e.\ $k$ quarks and antiquarks)
partitioned into $q\bar{q}$ singlets is seen to hold:

\begin{enumerate}
\item Draw two horizontal lines, the lower denoting time $-t/2$, the upper
$t/2$. Mark the position of every quark and antiquark on the lower line and
once again vertically above it on the upper line.
\item At the $-t/2$ level connect every quark-antiquark pair
that is set up as a singlet at $-t/2$ with a line,
having an arrow pointing from antiquark to quark.
\item At the $t/2$ level connect every quark-antiquark pair
that is set up as a singlet at $t/2$ with a line,
the arrow in which points from quark to antiquark.
\item Join the quarks at the $-t/2$ level
with quarks at the same position at
the $t/2$ level, arrow pointing upwards, i.e.\ forward in time.
\item Join the antiquarks at the $t/2$ level
with the antiquarks at the $-t/2$
level, arrow pointing downwards, i.e.\ backwards in time.
\item Associate a path-ordered exponential of
$e^{ig\oint_CT^aA_a^{\mu}(z)dz_{\mu}}$
together with a trace for every closed loop $C$ occurring.
\item Determine the overall sign:
If the pairings at the $-t/2$ level are the
same as those on the $t/2$ level, there must be a $+$ sign.
(This follows from the positivity of the norm on a Hilbert space
if one lets $t\rightarrow 0$.)
If this is not so, determine the sign of the permutation of antiquarks
on the upper line that is necessary to give the same pairings
as on the lower line. This is the overall sign.
\item Multiply by $(\delta(\vec{0})e^{-imt})^{k}$,
where $k$ is the total number
of quarks and antiquarks.
\item Insert the factor so obtained in the numerator\footnote{With $\eta$ we
denote the ghost fields, with $\cal L$ the Lagrangian without fermions} of
$\frac{\int[DA^a_{\mu}][D\eta_a^*][D\eta_a]e^{i\!\int\!d^4\!x[{\cal
L}]}}{\int[DA^a_{\mu}][D\eta_a^*][D\eta_a]e^{i\!\int\!d^4\!x[{\cal L}]}}$
\end{enumerate}
This gives the Green function in the chosen singlet structure.

\section{The $qq\bar{q}\bar{q}$-Potentials}
In $SU($N$)$ gauge theory with quarks in the fundamental representation,
we want to calculate the $qq\bar{q}\bar{q}$-potential
in perturbation theory to fourth order.
It has been remarked in subsection~\ref{sin} that there are two independent
basis states for
this system,
and one easily recognises a choice of these
in the two possible ways of pairing the system
into two quark-antiquark singlets.
Assuming the first static quark at position $R_1$, the second at $R_2$, and
the antiquarks at $R_3$ and $R_4$,
we will label the two states $|A_1\rangle =
{\bf 1}_{13}{\bf 1}_{24}$ and $|A_2\rangle =
{\bf 1}_{14}{\bf 1}_{23}$.

\subsection{Calculating the Green Functions}
According to subsection~\ref{loops},
we encounter the following types of loops:
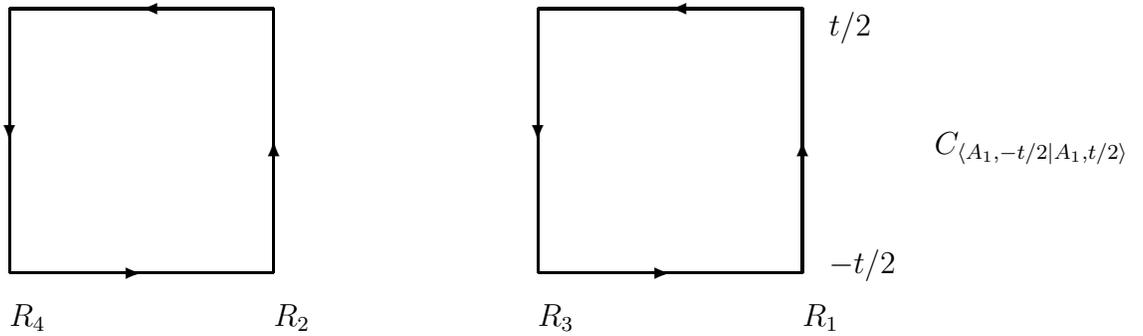
\begin{center}
\begin{picture}(350,150)
\thicklines
\put(0,35){\line(1,0){100}}
\put(0,35){\vector(1,0){ 50}}
\put(100,135){\line(-1,0){100}}
\put(100,135){\vector(-1,0){50}}
\put(300,135){\line(-1,0){100}}
\put(300,135){\vector(-1,0){50}}
\put(200,35){\line(1,0){100}}
\put(200,35){\vector(1,0){50}}
\put(0,135){\vector(0,-1){50}}
\put(0,135){\line(0,-1){100}}
\put(100,35){\vector(0,1){50}}
\put(100,35){\line(0,1){100}}
\put(200,135){\vector(0,-1){50}}
\put(200,135){\line(0,-1){100}}
\put(300,35){\line(0,1){100}}
\put(300,35){\vector(0,1){50}}
\put(  0, 15){$R_4$}
\put(100, 15){$R_2$}
\put(200, 15){$R_3$}
\put(300, 15){$R_1$}
\put(310, 35){$-t/2$}
\put(310,125){$t/2$}
\put(350,80){$C_{\langle A_1,-t/2|A_1,t/2\rangle}$}
\end{picture}
\end{center}
and
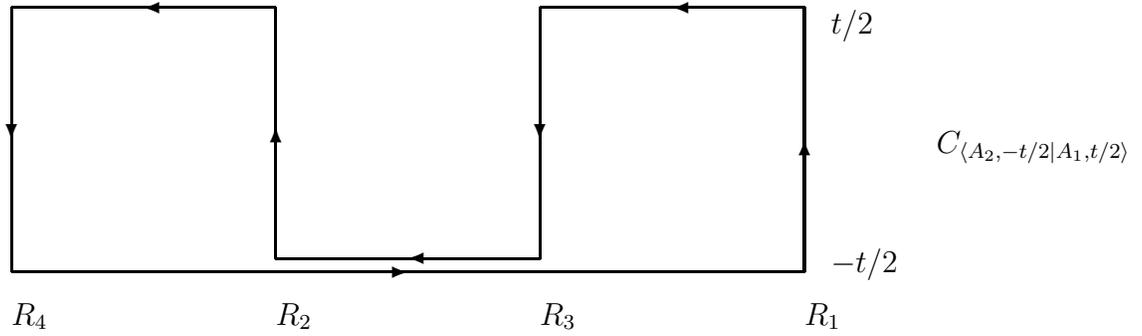
\begin{center}
\begin{picture}(350,150)
\thicklines
\put(0,35){\line(1,0){300}}
\put(0,35){\vector(1,0){150}}
\put(100,135){\line(-1,0){100}}
\put(100,135){\vector(-1,0){50}}
\put(300,135){\line(-1,0){100}}
\put(300,135){\vector(-1,0){50}}
\put(200,40){\line(-1,0){100}}
\put(200,40){\vector(-1,0){50}}
\put(0,135){\vector(0,-1){50}}
\put(0,135){\line(0,-1){100}}
\put(100,40){\vector(0,1){50}}
\put(100,40){\line(0,1){95}}
\put(200,135){\vector(0,-1){50}}
\put(200,135){\line(0,-1){95}}
\put(300,35){\line(0,1){100}}
\put(300,35){\vector(0,1){50}}
\put(  0, 15){$R_4$}
\put(100, 15){$R_2$}
\put(200, 15){$R_3$}
\put(300, 15){$R_1$}
\put(310, 35){$-t/2$}
\put(310,125){$t/2$}
\put(350,80){$C_{\langle A_2,-t/2|A_1,t/2\rangle}$}
\end{picture}
\end{center}

In calculating the Green functions,
we will adopt dimensional regularization in dimension $D=4-2\epsilon$
with a mass scale $M$.
A wave-function renormalization
will remove infinities associated with a diagram of the form
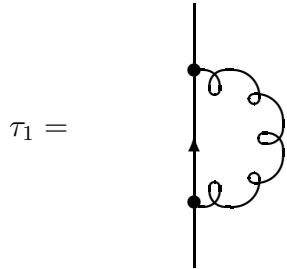
\input FEYNMAN
\begin{center}
\begin{picture}(10000,13000)
\thicklines
\put(5000,1500){\line(0,1){10000}}
\put(5000,1500){\vector(0,1){5000}}
\put(5000,4000){\circle*{500}}
\put(5000,9000){\circle*{500}}
\put(-2000,6500){${\tau}_1=$}
\drawline\gluon[\E\REG](5200,9000)[1]
\drawline\gluon[\SE\REG](\gluonbackx,\gluonbacky)[1]
\drawline\gluon[\S\REG](\gluonbackx,\gluonbacky)[1]
\drawline\gluon[\SW\REG](\gluonbackx,\gluonbacky)[1]
\drawline\gluon[\W\REG](\gluonbackx,\gluonbacky)[1]
\end{picture}
\end{center}
while a coupling constant renormalization is needed to make the sum
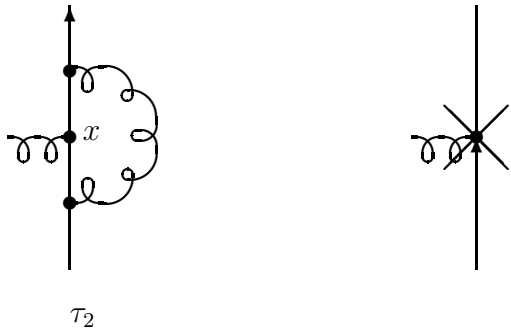
\begin{center}
\begin{picture}(10000,15000)
\thicklines
\put(5000,3500){\vector(0,1){10000}}
\put(5000,8500){\circle*{500}}
\put(5000,11000){\circle*{500}}
\put(5000,6000){\circle*{500}}
\put(5000,1500){${\tau}_2$}
\put(5500,8400){$x$}
\drawline\gluon[\E\FLIPPED](5100,6000)[1]
\drawline\gluon[\NE\FLIPPED](\gluonbackx,\gluonbacky)[1]
\drawline\gluon[\N\FLIPPED](\gluonbackx,\gluonbacky)[1]
\drawline\gluon[\NW\FLIPPED](\gluonbackx,\gluonbacky)[1]
\drawline\gluon[\W\FLIPPED](\gluonbackx,\gluonbacky)[1]
\drawline\gluon[\W\FLIPPED](4900,8500)[2]
\end{picture}
\begin{picture}(10000,13000)
\thicklines
\put(10000, 3500){\line(0,1){10000}}
\put(10000, 3500){\vector(0,1){ 5000}}
\put(10000,8500){\circle*{500}}
\put(10000,8500){\line(1,1){1200}}
\put(10000,8500){\line(1,-1){1200}}
\put(10000,8500){\line(-1,1){1200}}
\put(10000,8500){\line(-1,-1){1200}}
\drawline\gluon[\W\FLIPPED]( 9800,8500)[2]
\end{picture}
\end{center}
finite. The full gluon propagator is renormalized in the usual way.
Note that, contrary to a smallest distance regularization
frequently encountered in this context,
there is no renormalization of the quark mass $m$.
Making the expansion in $\alpha = \frac{g^2{\it c_3}}{4\pi^2}$
rather than in $g$ and also expanding
in time, a calculation yields for
\begin{equation}
{\bf A}={\bf A}\!(1)
+\alpha{\bf A}\!(\alpha)+\alpha^2{\bf A}\!(\alpha^2)+\alpha
t{\bf A}\!(\alpha t)+
\alpha^2t{\bf A}\!(\alpha^2t)+\frac{(\alpha t)^2}{2}{\bf A}\!
(\frac{(\alpha t)^2}{2})
\label{eq36}
\end{equation}
the expressions:
{\footnotesize
\begin{eqnarray}
\lefteqn{{\bf A}(1)=
   \left(
      \begin{array}{rr}
         N^2&-N\\-N&N^2
      \end{array}
   \right)}
\nonumber\\
&&\nonumber\\
\lefteqn{{\bf A}(\alpha)=
   \left(\!\!
      \begin{array}{ll}
         N^2 \left\{4 \ln(M^2R_{13}R_{24})\right\}&
         -N\left\{\ln\left(\frac{R^3_{13}R^3_{14}R^3_{23}
R^3_{24}M^8}{R^2_{12}R^2_{34}}
\right)-\mbox{e}_1-\mbox{e}_2\right\}\\
         -N\left\{\ln\left(\frac{R^3_{13}R^3_{14}R^3_{23}
R^3_{24}M^8}{R^2_{12}R^2_{34}}
\right)-\mbox{e}_1-\mbox{e}_2\right\}&
         N^2 \left\{4 \ln(M^2R_{14}R_{23})\right\}
      \end{array}
   \!\!\right)}
\nonumber\\
&&\nonumber\\
\lefteqn{{\bf A}(\alpha T)\,=
   \left(
      \begin{array}{ll}
         N^2\left\{-V_{s1}\right\}&
         -N\left\{-V_{s1}-V_{s2}+V_{d}\right\}\\
         -N\left\{-V_{s1}-V_{s2}+V_{d}\right\}&
         N^2\left\{-V_{s2}\right\}
      \end{array}
   \right)}
\nonumber\\
&&\nonumber\\
&&\nonumber\\
\lefteqn{{\bf A}(\alpha^2 T)\,=}
\nonumber\\
&&\nonumber\\
&&
   \left(
      \begin{array}{ll}
         \!\!N^2\left\{
                     -4V_{s1}\ln\left(M^2R_{13}R_{24}\right)
            \right.
                   &
         -N \left\{
                     \left[-V_{s1}-V_{s2}+V_{d}\right]\times
            \right.
         \\
            \left.
                     \mbox{\ \ \ \ }+\frac{2{\it c_2}}{N{\it
c_3}}\left[-V_{s2}+V_d\right]\times
            \right.
                   &
            \left.
                     \mbox{\ \ \ \
}\left[\ln\left(\frac{R^3_{13}R^3_{14}R^3_{23}R^3_{24}M^8}
{R^2_{12}R^2_{34}}\right)
                                          -\mbox{e}_1-\mbox{e}_2\right]
            \right.
         \\
            \left.
                     \mbox{\ \ \ \
}\left[\ln\left(\frac{R_{14}R_{23}}{R_{12}R_{34}}\right)-\mbox{e}_1\right]
            \right\}
                   &
            \left.
                     \mbox{\ \ \ \ }+\frac{{\it c_1}}{2{\it
c_3}}\left[\mbox{e}_1(V_d\!-\!V_{s1})

+\mbox{e}_2(V_d\!-\!V_{s2})\right]\!\!
            \right.
          \\
                   &
            \left.
                     \mbox{\ \ \ \ }-\frac{{\it c_1}}{2{\it
c_3}}\left[\ln(R_{12}R_{34})(V_{s1}\!+\!V_{s2}\!-\!2V_d)\right]
            \right.\!\!
          \\
                   &
            \left.
                     \mbox{\ \ \ \ }+\frac{{\it c_1}}{2{\it
c_3}}\left[\ln(R_{13}R_{24})(V_{s2}-V_d)\right]
            \right.
          \\
                   &
            \left.
                     \mbox{\ \ \ \ }+\frac{{\it c_1}}{2{\it
c_3}}\left[\ln(R_{14}R_{23})(V_{s1}-V_d)\right]
            \right\}
         \\
         \\
          \!\!-N\left\{
                     \left[-V_{s1}-V_{s2}+V_{d}\right]\times
            \right.
                   &
          N^2\left\{
                      -4V_{s2}\ln\left(M^2R_{14}R_{23}\right)
            \right.
         \\
            \left.
                     \mbox{\ \ \ \
}\left[\ln\left(\frac{R^3_{13}R^3_{14}R^3_{23}R^3_{24}M^8}
{R^2_{12}R^2_{34}}
\right)
                                          -\mbox{e}_1
-\mbox{e}_2\right]
            \right.
                   &
            \left.
                      \mbox{\ \ \ \ }+\frac{2{\it c_2}}{N{\it
c_3}}\left[-V_{s1}+V_d\right]\times
            \right.
          \\
            \left.
                     \mbox{\ \ \ \ }+\frac{{\it c_1}}{2{\it
c_3}}\left[\mbox{e}_1(V_d-V_{s1})

+\mbox{e}_2(V_d-V_{s2})\right]
            \right.
                   &
            \left.
                      \mbox{\ \ \ \
}\left[\ln\left(\frac{R_{13}R_{24}}{R_{12}R_{34}}\right)
-\mbox{e}_2\right]
            \right\}
          \\
            \left.
                      \mbox{\ \ \ \ }-\frac{{\it c_1}}{2{\it
c_3}}\left[\ln(R_{12}R_{34})(V_{s1}\!+\!V_{s2}\!-\!2V_d)\right]
            \right.
                   &
          \\
            \left.
                      \mbox{\ \ \ \ }+\frac{{\it c_1}}{2{\it
c_3}}\left[\ln(R_{13}R_{24})(V_{s2}-V_d)\right]
            \right.
                   &
          \\
            \left.
                      \mbox{\ \ \ \ }+\frac{{\it c_1}}{2{\it
c_3}}\left[\ln(R_{14}R_{23})(V_{s1}-V_d)\right]
            \right\}
                   &
      \end{array}
   \!\!\!\right)
\nonumber\\
&&\nonumber\\
&&\nonumber\\
&&\nonumber\\
\lefteqn{{\bf A}\left(\frac{1}{2}(\alpha T)^2\right)\,=}
\nonumber\\
&&\nonumber\\
&&
   \left(
      \begin{array}{ll}
         N^2\left\{V_{s1}^2+\frac{{\it c_2}}{N{\it c_3}}(V_{s2}-V_d)^2

            \right\}
                   &
         -N \left\{
                     (V_{s1}+V_{s2}-V_d)^2
                     +\frac{{\it c_1}}{2{\it c_3}}\times
            \right.
         \\
            \left.
                     \mbox{\ \ \ \ }
            \right.
                   &
            \left.
                     \mbox{\ \ \ \
}\left[V_dV_{s1}+V_dV_{s2}-V_{s1}V_{s2}-V_d^2\right]
            \right\}
         \\ \\
          -N\left\{
                     (V_{s1}+V_{s2}-V_d)^2
                     +\frac{{\it c_1}}{2{\it c_3}}\times
            \right.
                   &
          N^2\left\{V_{s2}^2+\frac{{\it c_2}}{N{\it
c_3}}\left(V_{s1}-V_d\right)^2

            \right\}
         \\
            \left.
                      \mbox{\ \ \ \
}\left[V_dV_{s1}+V_dV_{s2}-V_{s1}V_{s2}-V_d^2\right]
            \right\}
                   &
            \left.
                      \mbox{\ \ \ \ }
            \right.
      \end{array}
   \right)
\nonumber\end{eqnarray}
}
where, to save writing, the following shorthand notation has been adopted:
\begin{eqnarray*}
\mbox{edge}(\vec{R_3},\vec{R_1},\vec{R_4},\vec{R_2})\!\!
&\!\stackrel{def}{=}\!&
\!\!\int\!\!\int_{-1/2}^{1/2}
\frac{(\vec{R_3}-\vec{R_1})(\vec{R_4}-\vec{R_2})\,\,\,dw\,dx}
     {\left[\frac{\vec{R_3}+\vec{R_1}}{2}\!+\!w(\vec{R_3}\!-\!\vec{R_1})
           \!-\!\frac{\vec{R_4}\!+\!\vec{R_2}}{2}\!-\!x(\vec{R_4}\!
           -\!\vec{R_2})
      \right]^2}\\
\mbox{e}_1&\stackrel{def}{=}&\mbox{edge}(R_1,R_3,R_2,R_4)\\
\mbox{e}_2&\stackrel{def}{=}&\mbox{edge}(R_1,R_4,R_2,R_3)\\
V_{s1}&\stackrel{def}{=}&V(R_{13})+V(R_{24})\\
V_{s2}&\stackrel{def}{=}&V(R_{14})+V(R_{23})\\
V_{d}&\stackrel{def}{=}&V(R_{12})+V(R_{34})\\
\end{eqnarray*}
\vspace{-5 ex}
\[
\begin{array}{rrrrr}
{\it c_1}\delta_{ab}&=&\sum_{cd}{\it f}_{acd}{\it f}_{bcd}&=&N\delta_{ab}
\\
{\it c_2}\delta_{ab}&=&{\it Tr\/}[T^aT^b]&=&\frac{1}{2}\delta_{ab}
\\
{\it c_3}{\bf 1}&=&\sum_a T^aT^a&=&\frac{N^2-1}{2N}{\bf 1}
\end{array}
\]
and $V(R_{pq})$ is the two-body potential
between a quark `p' and an antiquark
`q' a
distance $R_{pq}$ apart.
Diagonalization yields for the two possible energy eigenstates
the two potentials correct to fourth order:
\begin{eqnarray}
V_0\!&=&\!\frac{
\left(N^2\!-\!2\right)
\left(V_{s1}\!+\!V_{s2}\right)
+2V_{d}
-N\sqrt{
N^2\left(V_{s1}\!-\!V_{s2}\right)^2
+4\left(V_{s1}\!-\!V_d\right)\!\left(V_{s2}\!
-\!V_d\right)
}
}{2\left(N^2-1\right)}
\nonumber\\
V_1\!&=&\!\frac{
\left(N^2\!-\!2\right)
\left(V_{s1}\!+\!V_{s2}\right)
+2V_{d}
+N\sqrt{
N^2\left(V_{s1}\!-\!V_{s2}\right)^2
+4\left(V_{s1}\!-\!V_d\right)\!\left(V_{s2}\!
-\!V_d\right)
}
}{2\left(N^2-1\right)}
\nonumber\\
\label{potentials}
\end{eqnarray}
These potentials are exactly equal to those
given by a naive two-body model (see~\cite{paton})---and are one of the
main conclusions of this work.

\subsection{Consistency Relations and Edge Effects}
When integrating out the gluonic degrees of freedom
and reducing a field theory to a quantum mechanical system,
we have made assumptions which may need some justification.
This is especially so in the perturbative case where we have postulated
a finite number of energy eigenstates.
We, therefore, want to have a look at the consistency of this approach.

There are certain
relations which must hold in order to guarantee proper exponentiation.
We have remarked at the end of subsection~\ref{diag}
that after diagonalization we expect the diagonal
entries of the matrix $\bf A$ to be
(apart from a normalization factor) of the form $e^{-itE_i(g)}$.
An actual calculation will give a power expansion in $g$ and $t$ for
these diagonal entries. This will not only determine the energy $E_i(g)$
as the coefficient of $t$, but will also allow us
to check the consistency of our calculation by inspecting
the relations between the coefficients of higher powers of $t$.
We were able to verify these consistency relations
in our calculation to fourth order.

Another important point to notice is that the terms we have abbreviated
$\mbox{edge}(\vec{R_p},\vec{R_q},\vec{R_r},\vec{R_s})$ have cancelled in
eq.~(\ref{potentials}).
These terms are coming from the space-like contour integrations. In
subsection~\ref{sin}
we had introduced these space-like contour integrations
via the path-ordered exponential $U$ in order to establish
colour singlets. The exact integration contour in $U$
was of course arbitrary, and the physical potentials
must not depend on it. Another way of looking at it is that
in the large time limit the processes associated with bringing the
quarks into their position in the distant past
(and removing them again in the distant future)
should become irrelevant.
One may also make an adiabatic argument, turning on the coupling
$g$ in the distant past and off again in the distant future,
thus making the notion of a colour singlet state at these times
well-defined in spite of the spatial separation of the quarks.
In any case, for the potentials of eq.~(\ref{potentials}) to be
meaningful, they must not involve terms originating in contour
integrations in the distant past or future, and indeed they do not.

\subsection{Three- and Four-Body Forces in Sixth Order}
The fact that a straightforward two-body model is correct also to
next-to-leading order
may be surprising in light of the non-abelian nature of QCD.
Hence we want to mention that we believe that the two-body model fails
at sixth order as three- and four-body forces appear at this order.
While many diagrams that seem to give rise to such deviations
from the two-body model actually vanish, we see no reason why
for example effects from the following two diagrams should cancel
in the calculation of the four-quark potentials to sixth order.
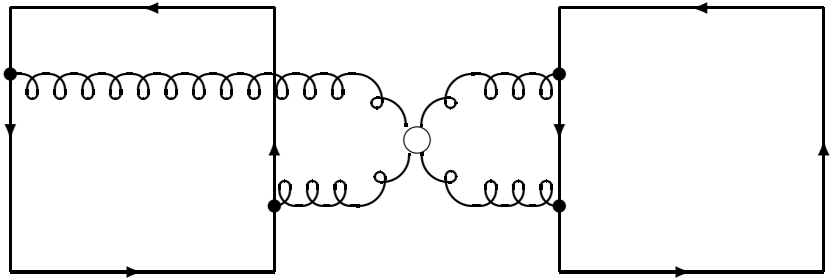
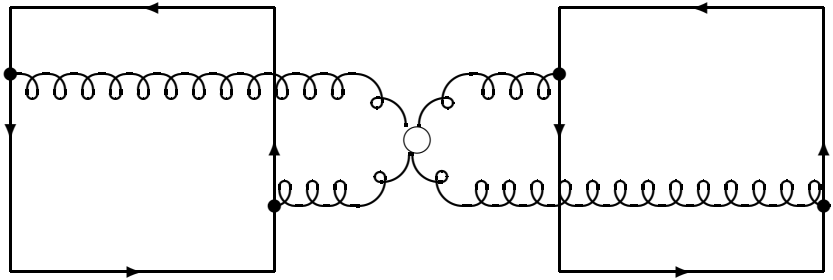
\begin{center}
\begin{picture}(10000,13000)
\thicklines
\put(0,11500){\vector(0,-1){5000}}
\put(10000,1500){\vector(0,1){5000}}
\put(0,1500){\vector(1,0){5000}}
\put(10000,11500){\vector(-1,0){5000}}
\put(0,11500){\line(0,-1){10000}}
\put(10000,1500){\line(0,1){10000}}
\put(0,1500){\line(1,0){10000}}
\put(10000,11500){\line(-1,0){10000}}
\put(0,9000){\circle*{500}}
\put(10000,4000){\circle*{500}}
\drawline\gluon[ \E    \REG](  250, 9000)[12]
\drawline\gluon[\SE    \REG](\gluonbackx,\gluonbacky)[1]
\drawline\gluon[ \E\FLIPPED]( 9800, 4000)[3]
\drawline\gluon[\NE\FLIPPED](\gluonbackx,\gluonbacky)[1]
\end{picture}
\begin{picture}(10000,13000)
\put(5000,6500){\circle{1000}}
\end{picture}
\begin{picture}(10000,13000)
\thicklines
\put(0,11500){\vector(0,-1){5000}}
\put(10000,1500){\vector(0,1){5000}}
\put(0,1500){\vector(1,0){5000}}
\put(10000,11500){\vector(-1,0){5000}}
\put(0,11500){\line(0,-1){10000}}
\put(10000,1500){\line(0,1){10000}}
\put(0,1500){\line(1,0){10000}}
\put(10000,11500){\line(-1,0){10000}}
\put(0,9000){\circle*{500}}
\put(0,4000){\circle*{500}}
\drawline\gluon[ \W\FLIPPED](  100, 9000)[3]
\drawline\gluon[\SW\FLIPPED](\gluonbackx,\gluonbacky)[1]
\drawline\gluon[ \W    \REG](  100, 4000)[3]
\drawline\gluon[\NW    \REG](\gluonbackx,\gluonbacky)[1]
\end{picture}
\end{center}
\begin{center}
\begin{picture}(10000,15000)
\thicklines
\put(0,13500){\vector(0,-1){5000}}
\put(10000,3500){\vector(0,1){5000}}
\put(0,3500){\vector(1,0){5000}}
\put(10000,13500){\vector(-1,0){5000}}
\put(0,13500){\line(0,-1){10000}}
\put(10000,3500){\line(0,1){10000}}
\put(0,3500){\line(1,0){10000}}
\put(10000,13500){\line(-1,0){10000}}
\put(0,11000){\circle*{500}}
\put(10000,6000){\circle*{500}}
\drawline\gluon[ \E    \REG](  250,11000)[12]
\drawline\gluon[\SE    \REG](\gluonbackx,\gluonbacky)[1]
\drawline\gluon[ \E\FLIPPED]( 9800, 6000)[3]
\drawline\gluon[\NE\FLIPPED](\gluonbackx,\gluonbacky)[1]
\end{picture}
\begin{picture}(10000,15000)
\put(5000,8500){\circle{1000}}
\end{picture}
\begin{picture}(10000,15000)
\thicklines
\put(0,13500){\vector(0,-1){5000}}
\put(10000,3500){\vector(0,1){5000}}
\put(0,3500){\vector(1,0){5000}}
\put(10000,13500){\vector(-1,0){5000}}
\put(0,13500){\line(0,-1){10000}}
\put(10000,3500){\line(0,1){10000}}
\put(0,3500){\line(1,0){10000}}
\put(10000,13500){\line(-1,0){10000}}
\put(0,11000){\circle*{500}}
\put(10000,6000){\circle*{500}}
\drawline\gluon[ \W\FLIPPED](    0,11000)[3]
\drawline\gluon[\SW\FLIPPED](\gluonbackx,\gluonbacky)[1]
\drawline\gluon[ \W    \REG](10250, 6000)[13]
\drawline\gluon[\NW    \REG](\gluonbackx,\gluonbacky)[1]
\end{picture}
\end{center}
\begin{center}
\begin{picture}( 3000, 8000)
\put(    0, 4000){where}
\end{picture}
\begin{picture}( 5000, 8000)
\put( 2500, 4000){\circle{1000}}
\end{picture}
\begin{picture}( 5000, 8000)
\put(    0, 4000){stands for}
\end{picture}
\begin{picture}(15000,8000)
\thicklines
\put(5000,4000){\circle*{500}}
\put(10000,4000){\circle*{500}}
\drawline\gluon[\E \REG    ]( 5250, 4000)[4]
\drawline\gluon[\NW\FLIPPED]( 5000, 4000)[2]
\drawline\gluon[\SW\FLIPPED]( 5000, 4000)[2]
\drawline\gluon[\NE\REG    ](10000, 4000)[2]
\drawline\gluon[\SE\REG    ](10000, 4000)[2]
\end{picture}
\end{center}
Hence three- and four-body forces will be introduced.
In general, their nature seems to be complicated, but
for some geometries simplifications are possible; e.g.
for the four quarks on the corners of a regular
tetrahedron there will be no contribution from quark self-
interactions to four-body forces to sixth order.

\section{Relation of our Result to Lattice Simulations}

Looking at the Monte Carlo lattice calculations for the
$qq\bar{q}\bar{q}$-system
discussed in~\cite{paton}, we observe that for small interquark distances
of a few lattice spacings (with a lattice spacing of $a\approx 0.12$ fm)
the two-body model gives a reasonable approximation in the sense
that the four-quark potentials calculated from eq.~(\ref{potentials})
using the Monte Carlo two-body potentials are comparable to the four-quark
potentials from the lattice simulation.
The agreement improves the smaller the distances get. By comparing the
perturbative (i.e.\ $1/R$) and
non-perturbative (i.e. linear) part in the usual parametrization of the
$q\bar{q}$-potential,
one would expect to start entering the perturbative regime
at distances of about two lattice spacings.
However, at that stage the approximation
provided by the two-body model is already very good. The
fact that the two-body model is correct to fourth order
in perturbation theory
certainly suggests that it should be a reasonable approximation
in the perturbative domain. So our result supports the belief
that the results of the lattice simulations for small
enough distances indeed are correlated
to continuum perturbation theory,
and thus that continuum physics is extracted
from the Monte Carlo calculations.

\vspace{5 ex}
\paragraph{Acknowledgements}
One of the authors (J. L.) wants to express his gratitude to the `Stiftung
Maximilianeum' which made possible his stay at Oxford,
during which time most of
this work was carried out.

\end{document}